\documentclass[11pt]{article}
\usepackage{amsmath,amssymb,color,epsfig,cite}
\usepackage{xcolor}
\usepackage{mathrsfs}

\usepackage{mathtools}

\usepackage{amsfonts}

\newcommand{\hoch}[1]{$\, ^{#1}$}

\makeatletter
\@addtoreset{equation}{section}
\makeatother

\newcommand{\be}{\begin{equation}}
\newcommand{\ee}{\end{equation}}
\newcommand{\bea} {\begin{eqnarray}}
\newcommand{\eea}{\end{eqnarray}}
\newcommand{\nn}{\nonumber}

\def\ft#1#2{{\textstyle{\frac{\scriptstyle #1}{\scriptstyle #2} } }}
\def\fft#1#2{{\frac{#1}{#2}}}

\def\0{{\sst{(0)}}}
\def\1{{\sst{(1)}}}
\def\2{{\sst{(2)}}}
\def\3{{\sst{(3)}}}
\def\4{{\sst{(4)}}}
\def\5{{\sst{(5)}}}
\def\6{{\sst{(6)}}}
\def\7{{\sst{(7)}}}
\def\8{{\sst{(8)}}}
\def\sst#1{{\scriptscriptstyle #1}}
\def\oneone{\rlap 1\mkern4mu{\rm l}}

\def\del{{\partial}}

\def\crampest{\medmuskip = 1mu plus 1mu minus 1mu}
\def\uncramp{\medmuskip = 4mu plus 2mu minus 4mu}

\def\cG{{{\cal G}}}

\def\im{{{\rm i\,}}}
\def\R{{\mathbb R}}

\begin{document}

\begin{flushright}
\hfill { UPR-1302-T\ \ \ MI-TH-2022
}\\
\end{flushright}

\begin{center}
{\large {\bf Generalised Couch-Torrence Symmetry for 
Rotating Extremal Black Holes in Maximal Supergravity}}

\vspace{15pt}
{\large M. Cveti\v c$^{1,2}$, 
              C.N. Pope$^{3,4}$ and A. Saha$^{3}$}

\vspace{15pt}

{\hoch{1}\it Department of Physics and Astronomy,\\
University of Pennsylvania, Philadelphia, PA 19104, USA}

\vspace{10pt}

{\hoch{2}\it Center for Applied Mathematics and Theoretical Physics,\\
University of Maribor, SI2000 Maribor, Slovenia}

\vspace{10pt}

\hoch{3}{\it George P. \& Cynthia Woods Mitchell  Institute
for Fundamental Physics and Astronomy,\\
Texas A\&M University, College Station, TX 77843, USA}

\hoch{4}{\it DAMTP, Centre for Mathematical Sciences,
 Cambridge University,\\  Wilberforce Road, Cambridge CB3 OWA, UK}

\vspace{10pt}



\end{center}




\begin{abstract}

  The extremal Reissner-Nordstr\"om black hole admits a conformal inversion
symmetry, in which the metric is mapped into itself under an inversion of the 
radial coordinate combined with a conformal rescaling. In the rotating 
generalisation, Couch and Torrence showed that 
the Kerr-Newman metric no longer exhibits a conformal
inversion symmetry, but the radial equation arising in the separation of 
the massless Klein-Gordon equation admits a mode-dependent 
inversion symmetry, where the
radius of inversion depends upon the energy and azimuthal 
angular momentum of the mode.  It was more recently shown that the
static 4-charge extremal black holes of STU supergravity admit a 
generalisation of the conformal inversion symmetry, in which the 
conformally-inverted metric is a member of the same 4-charge black hole
family but with transformed charges.  In this paper we study further
generalisations of these inversion symmetries, 
 within the general class of extremal STU supergravity black holes.  For
the rotating black holes, where again the massless Klein-Gordon equation is
separable, we show that examples with four electric charges exhibit a 
generalisation of the Couch-Torrence symmetry of the radial equation.  Now,
as in the conformal inversion of the static specialisations, the 
inversion of the radial equation maps it to the radial equation for a
rotating black hole with transformed electric charges.  We also study
the inversion transformations for the general case of extremal BPS STU black
holes carrying eight charges (4 electric plus 4 magnetic), and argue that
analogous generalisations of the inversion symmetries exist both for 
the static and the rotating cases.

\end{abstract}

\pagebreak

\tableofcontents
\addtocontents{toc}{\protect\setcounter{tocdepth}{2}}



\def\half{\frac{1}{2}}
\def\ben{\begin{equation}}
\def\bea{\begin{eqnarray}}
\def\een{\end{equation}}
\def\eea{\end{eqnarray}}
\def \bp {{\bf p}}
\def \bv {{\bf v}}
\def \bs {{\bf s}}
\def\bt{{\bf t}}

\def \p {\partial}
\def \cL {{\cal  L}}
\def \cG {{\cal  G}}
\def \cLEG {{\cal LEG}}

\def\ft#1#2{{\textstyle{\frac{\scriptstyle #1}{\scriptstyle #2} } }}
\def\fft#1#2{{\frac{#1}{#2}}}

\section{Introduction}

   It was observed many years ago that the extremal limit of the 
Reissner-Nordstr\"om black hole exhibits a remarkable conformal inversion
symmetry, in which an inversion of the radial coordinate, which maps the
near-horizon region to the region near infinity, combined with a
conformal rescaling, transforms the original metric back into itself
\cite{coutor}.  Explicitly, consider the original extremal Reissner-Nordstr\"om
metric written in the isotropic form
\be
ds^2= -\Big(1+\fft{Q}{r}\Big)^{-2}\, dt^2 + \Big(1+\fft{Q}{r}\Big)^2\,
(dr^2 + r^2\, d\Omega_2^2)\,,\label{RNmet}
\ee
where $d\Omega_2^2$ is the metric on the unit 2-sphere.  The horizon is
located at $r=0$.  Performing the inversion to a new radial coordinate
\be
\tilde r= \fft{Q^2}{r}\,,\label{radiusinv}
\ee
one finds that the conformally-related metric $d\tilde s^2$ defined by
\be
ds^2 = \fft{Q^2}{\tilde r^2}\, d\tilde s^2
\ee
is given by
\be
d\tilde s^2= -\Big(1+\fft{Q}{\tilde r}\Big)^{-2}\, dt^2 
  + \Big(1+\fft{Q}{\tilde r}\Big)^2\,
(d\tilde r^2 + \tilde r^2\, d\Omega_2^2)\,.\label{tRNmet}
\ee
Thus after the conformal inversion, the resulting metric $d\tilde s^2$ is
again the extremal Reissner-Nordstr\"om metric, and takes the identical
form to the original metric (\ref{RNmet}) \cite{coutor}.  The radius 
of the inversion in (\ref{radiusinv}) is equal to the electric charge
$Q$ (and hence also the mass).

   The conformal inversion symmetry of the extremal Reissner-Nordstr\"om
metric has been employed in a number of papers (see, for example, 
\cite{bizfri,lumureta,godgodpop}) in order to relate some
of the recent observations about the asymptotic behaviour of 
Klein-Gordon and other fields on the future horizon of an extremal
black hole (see, for
example, \cite{aretakis1,aretakis2}) to the
asymptotic behaviour of these fields at future null infinity.   

   It was shown also in \cite{coutor} that the conformal inversion
symmetry of the extremal Reissner-Nordstr\"om metric does not
generalise to the rotating case, namely the extremal Kerr-Newman
metric. It was, however, observed that if one considers a massless
Klein-Gordon field $\Phi$ in the extremal Kerr-Newman background, then after
performing a separation of variables the differential equation for the
radial function exhibits a remarkable inversion symmetry. This 
involves a 
radius of inversion that depends not only on the black hole charge and
rotation 
parameters but also on the separation constants $\omega$ and $m$ that
arise in the factorised solutions
\be
\Phi(t,r,\theta,\varphi)= R(r)\, S(\theta)\, e^{-\im\omega t}\, 
e^{\im m \varphi}\,.
\ee
Thus the inversion radius is different for different modes.

   In \cite{godgodpop}, conformal inversions of a class of more general
static extremal black holes were investigated.  Specifically, the
static extremal 4-charge black hole solutions \cite{cvyoII} of the four-dimensional
STU supergravity theory   were studied.  STU supergravity comprises
${\cal N}=2$ supergravity coupled to three vector 
supermultiplets.\footnote{This is also a consistent truncation of ungauged 
supergravity theory with maximal (both ${\cal N}=4$ and ${\cal N}=8$) 
supersymmetry. Black hole solutions of STU supergravity are generating 
solutions of the full maximally supersymmetric  ungauged supergravity 
theories.} 
Thus there are four electromagnetic field strengths in total, each of
which can carry, in general, independent electric and magnetic charges.
The general 8-charge solution is quite complicated but the special case
where each field strength carries just an electric charge is much simpler,
with the metric being given by
\bea
ds^2 &=& - H^{-1/2}\, dt^2 + H^{1/2}\, (dr^2+ r^2\, d\Omega_2^2)\,,
\label{4qmet}\\
H&=& \prod_{i=1}^4 \Big(1+ \fft{Q_i}{r}\Big)\,,
\eea
As was shown in \cite{godgodpop}, under the inversion
\be
\tilde r= \fft{\Pi^2 }{r}\,,\qquad \Pi\equiv \prod_{i=1}^4 Q_i^{1/4}\,,
\ee
one finds that the conformally related metric $d\tilde s^2$ given by
\be
ds^2= \fft{\Pi^2}{\tilde r^2}\, d\tilde s^2
\ee
takes the form
\bea
d\tilde s^2 &=& 
- \widetilde H^{-1/2}\, dt^2 + \widetilde H^{1/2}\, 
(d\tilde r^2+ \tilde r^2\, d\Omega_2^2)\,,
\label{4tqmet}\\
\widetilde H&=& \prod_{i=1}^4 \Big(1+ \fft{\widetilde Q_i}{\tilde r}\Big)\,,
\eea
where
\be
\widetilde Q_i = \fft{\Pi^2}{Q_i}\,.\label{tqq}
\ee
Thus the metric $d\tilde s^2$ obtained by the conformal inversion is
in the same class of 4-charge static extremal black holes as the
original metric (\ref{4qmet}), but for new charges $\widetilde Q_i$
related to the original charges $Q_i$ by (\ref{tqq}) \cite{godgodpop}. 

 In the special case where all four charges are equal, the metric (\ref{4qmet})
reduces to the extremal Reissner-Nordstr\"om  metric (\ref{RNmet}) and
the conformal inversion gives back this metric again.  A more general
specialisation where the conformal inversion becomes an actual 
symmetry is if the charges are set equal in pairs; for example
\be
Q_3=Q_1\,,\quad Q_4=Q_2\,,\qquad \implies \qquad \widetilde Q_3=
 \widetilde Q_1=Q_2\,,
\quad \widetilde Q_4=\widetilde Q_2 = Q_1
\ee
and hence $\widetilde H=H$ \cite{godgodpop}.

   One purpose of the present paper is to investigate whether the 
observation of Couch and Torrence, that the conformal inversion symmetry
of the extremal static Reissner-Nordstr\"om metric has a corresponding
inversion transformation of the radial equation in the rotating case,
might generalise to rotating versions of the 4-charge extremal black
holes in STU supergravity.  We show that this question can be answered
in the affirmative.  Namely, we show that in the rotating extremal
4-charge STU supergravity background, the radial equation for the
separated massless Klein-Gordon equation indeed 
exhibits an inversion symmetry.  As in the Kerr-Newman case, the inversion
radius depends not only on the black hole charge and rotation 
parameters but also on the mode numbers $\omega$ and $m$ 
arising in the separation of variables.  A new feature that we find for
the general 4-charge black holes is that after the inversion, the
radial equation is not the same as the original radial equation but,
rather, it is the radial equation for a transformed set of charge 
parameters.  This is the analogue, for the rotating case, of the mapping 
of the charges (\ref{tqq}) that was found for the static 4-charge black holes.

  Another purpose of this paper is to investigate 
conformal inversion in the most general setting of the
8-charge static extremal STU supergravity black holes, and the
analogous inversion of the Klein-Gordon radial equation in the 
extremal 8-charge rotating backgrounds.  The goal is to establish 
whether these inversion transformations again provide a mapping back into
the solution space of the 8-charge black holes.  

   It should be emphasised at this point that there exist disjoint classes
of static extremal black holes in STU supergravity.  The examples
with four electric charges that we discussed above are contained within
the class of BPS extremal black holes.  These are typically supersymmetric,
preserving some fraction of the supersymmetry that is governed by the
number, and the pattern, of the non-vanishing charges.  These BPS
black holes form the focus of our studies in this paper.  Our findings, 
which we describe in detail later, suggest that the entire
8-charge static BPS family of black holes maps into itself 
under conformal inversion.  We reach this conclusion by constructing
explicitly the general 8-charge static BPS extremal family of black 
holes, and studying the conditions that arise from requiring that the
family map into itself under conformal inversion.  Unlike the
situation for the case of four electric charges which was studied in
\cite{godgodpop}, it does not seem to be possible in the general
case to give an elegant formula for the mapping of the charges under
conformal inversion, analogous to that in eqn (\ref{tqq}).  This
is a consequence of the fact that the system of conditions resulting from
the requirement of the existence of a conformal inversion mapping is
underdetermined.  That is to say, there are now fewer conditions than the number
of unknowns (the eight mapped charges), and so there is not a unique
solution.  We have checked in many examples, and it appears that a
mapping of charges always exists.

   We also study the general 8-charge rotating extremal black holes.
As in the 4-charge specialisation described earlier, here too the massless
Klein-Gordon equation can be separated and the behaviour of the radial
wave equation under inversion can be investigated.  We find that, as
in the 8-charge static BPS extremal black holes described above, although
we can write down the conditions for the inversion of the radial
coordinate to give rise again to a radial equation for a set of mapped
charges, it does not appear to be possible to give an elegant formula
for the mapped charges.  Again, the reason is that the system of
conditions for invertability does not fully constrain the mapped charges.
As in the static case, we may nevertheless argue that solutions for
the mapped charges will exist.

There also exists a class of extremal black holes in STU supergravity
that does not obey the BPS conditions.  A simple example that was
investigated in \cite{godgodpop} was the static extremal Kaluza-Klein dyonic
black hole, which in the language of STU supergravity corresponds to the
case where just one of the four electromagnetic fields is turned on
and carries independent electric and magnetic charges, with the
other three electromagnetic fields being zero.  It was shown in 
\cite{godgodpop} that this metric does not map into any metric within the
same family, under conformal inversion.  This does not, however, necessarily
provide a counter-example to the idea that the entire 8-charge family 
of non-BPS extremal black holes might map
into itself under conformal inversion, since the conformal inversion
of the Kaluza-Klein dyon might in principle be an 8-charge non-BPS extremal solution that
did not lie within the original Kaluza-Klein dyonic subset.  However, 
owing to the greater complexity of the general 8-charge non-BPS extremal
solutions, we shall not pursue this question further in the present paper.

\section{Inversion Symmetry of Radial Equation for 
4-Charge Rotating Black Holes}
\label{sec:4chargerot}

   Couch and Torrence observed that although there is no conformal
inversion symmetry of the extremal Kerr-Newman metric (unlike the situation
for the extremal Reissner-Nordstr\"om solution), there is nevertheless
an inversion symmetry of the radial equation after one separates variables
in the massless scalar wave equation in the Kerr-Newman geometry 
\cite{coutor}.  Here,
we show that there exists a generalisation of this inversion symmetry
for the radial equation arising from separation of variables for the
massless scalar wave equation in the four-dimensional
4-charge rotating black hole solutions.  The situation for the general
4-charge case is reminiscent of the situation found in \cite{godgodpop} 
for the entire metric of the 4-charge extremal static black holes, 
namely, that the inversion
applied to a generic 4-charge case maps it into another case with 
{\it different} values of the four charges.  In the special case of
pairwise-equal charges, the inversion maps the radial equation into
exactly the same radial equation.  A special case of this, when all
four charges are equal, reduces to the Kerr-Newman result found by
Couch and Torrence.

   A convenient presentation of the rotating black holes in four-dimensional
STU supergravity carrying four independent charges  can be found in
\cite{cvyo,chcvlupo}.  The metric can be written as \cite{chcvlupo}:
\bea
ds^2 &=& -\fft{\rho^2 -2 m r}{W}\, (dt + {\cal B}_\1 )^2 +W\, 
\Big(\fft{dr^2}{\Delta} + d\theta^2 + \fft{\Delta\, 
  \sin^2\theta\, d\tilde\phi^2}{\rho^2-2mr}\Big)\,,\label{4chargemet}\\
{\cal B}_\1 &=& \fft{2m a \sin^2\theta\, (r\, \Pi_c - (r-2m)\, \Pi_s)}{
\rho^2 -2m r}\, d\tilde\phi\,,\quad \rho^2=r^2 + a^2\, \cos^2\theta\,,
\quad \Delta= r^2 -2m r + a^2\,,\nn\\
W^2 &=& R_1 R_2 R_3 R_4 + a^4\, \cos^4\theta \nn\\
&&+
\Big[ 2r^2 + 2mr \sum_i s_i^2 + 8m^2\, (\Pi_c-\Pi_s)\, \Pi_s -
   4m^2\, \sum_{i<j<k} s_i^2\, s_j^2\, s_k^2\Big]\,a^2\, 
\cos^2\theta\,,\nn\\
R_i&=& r+ 2 m s_i^2\,,\qquad \Pi_c=\prod_i c_i\,,\qquad \Pi_s=\prod_i s_i\,.\nn
\eea
The four physical charges are given by $q_i=2m s_i\, c_i= m \sinh2\delta_i$.
Extremality is achieved by taking $m=a$, and the horizon is then at $r=a$. 
   It is straightforward to separate variables in the massless wave
equation $\square\Psi=0$ 
in this background.\footnote{The massless wave equation is separable also in the non-extremal case, $m\ne a$, as shown in \cite{cvla}.}  Writing
\be
\Psi= e^{-\im\omega t+\im m \tilde\phi}\,R(\hat r)\, S(y)\,,
\ee
where we write $y=\cos\theta$ and define
\be
\hat r= r-a
\ee
(so that the horizon is located at $\hat r=0$), we find that $R$ and $S$ 
satisfy  the equations
\be
\hat r^2\,R'' + 2\hat r\,R' + (H - \lambda)\,R=0
\,,\qquad
(1-y^2)\,S'' -2y\, S' -\Big( \fft{m^2}{1-y^2} +a^2\,\omega^2\, (1-y^2)-
 \lambda\Big)\, S=0\,,
\ee
where $\lambda$ is the separation constant and
\crampest
\bea
H &=& 
\omega^2\, \hat r^2 +2  a\omega^2\,\Big(2+\sum_i s_i^2\Big)\, 
\hat r +
 2 a^2\,\omega^2\, \Big(4 + 3\sum_i s_i^2 + 2\sum_{i<j} s_i^2\, s_j^2\Big)\nn\\
&&
+\Big[ 8 a^3\,\omega^2 \, \Big(1+ \sum_i s_i^2 + \sum_{i<j} s_i^2\, s_j^2
+\sum_{i<j<k} s_i^2\, s_j^2\, s_k^2\Big) -
4 a^2\,\omega m\, (\Pi_c-\Pi_s)\Big]\, \fft1{\hat r} \\
&&\!\!\!\!\!\!\!\! \!\!\!\!\!\!\!\! \!\!\!\!\!\!\!\! +
\Big[ a^2\,m^2 - 4 a^3\, \omega\,m\, (\Pi_c+\Pi_s) +
4 a^4\,\omega^2\, \Big(1+ \sum_i s_i^2 + \sum_{i<j} s_i^2\, s_j^2
+\sum_{i<j<k} s_i^2\, s_j^2\, s_k^2 +2\Pi_s\, (\Pi_c+\Pi_s)\Big)\Big]\,
\fft1{\hat r^2}\,.\nn
\eea
\uncramp
A straightforward calculation then shows that, writing $H$ as 
$H(\hat r,\delta_i)$, with $s_i=\sinh\delta_i$, etc., then
\be
H(\hat r,\delta_i)= H(\fft{\beta^2}{\hat r},\tilde\delta_i)\,,
\ee
where
\be
\beta^2 = -\fft{a m}{\omega} + 2a^2\,(\Pi_c+\Pi_s)\,,
\ee
and the redefined charge parameters $\tilde\delta_i$ are related to the 
original parameters $\delta_i$ by the involution $\tilde\delta_i=
\ft12(\delta_1+\delta_2+\delta_3+\delta_4)-\delta_i$, i.e.
\bea
\tilde\delta_1&=&\ft12(\delta_2+\delta_3+\delta_4-\delta_1)\,,
\qquad
\tilde\delta_2=\ft12(\delta_1+\delta_3+\delta_4-\delta_2)\,,\nn\\
\tilde\delta_3&=&\ft12(\delta_1+\delta_2+\delta_4-\delta_3)\,,
\qquad
\tilde\delta_4=\ft12(\delta_1+\delta_2+\delta_3-\delta_4)\,.\label{invol}
\eea

  Let us now define a (dimensionless) 
coordinate $x=\beta/\hat r$.  Viewing the radial
function $R$ as now being a function $P(x)$, we see that it satisfies
\be
x^2\, \del_x^2\, P(x) + f(x,\delta_i)\, P(x)=0\,,\label{Peqn}
\ee
where
\bea
f(x,\delta_i) &=& \fft1{x^2}\, \Big[ H(\fft{\beta}{x},\delta_i)-\lambda\Big]
\nn\,,\\
&=& \beta^2\, \omega^2 +\fft{\beta^2\, \omega^2}{x^4} +
   \fft{C_+(\delta_i)}{x} + \fft{C_-(\delta_i)}{x^3} + 
  \fft{C_0(\delta_i)-\lambda}{x^2}\,,
\eea
with
\bea
C_+(\delta_i) &=& 4 a \beta\omega^2\, (\Pi_c-\Pi_s)\,,\qquad
C_-(\delta_i) = a\beta\omega^2\, \sum_i\cosh2\delta_i\,,\nn\\
C_0(\delta_i)&=&
a^2\,\omega^2\,\Big[2 +
\fft12\sum_{i<j}(\cosh2(\delta_i-\delta_j)+\cosh2(\delta_i+\delta_j))\Big]\,.
\eea
One can easily verify that $\beta$ and $C_0$ are invariant under $\delta_i
\rightarrow \tilde\delta_i$, while
\be
C_+(\delta_i)=C_-(\tilde\delta_i)\,,\qquad
C_-(\delta_i) = C_+(\tilde\delta_i)\,.
\ee
It follows that
\be
f(\fft1{x},\delta_i)= x^4\, f(x,\tilde\delta_i)\,,
\ee
and so if $P(x,\delta_i)$ is a solution of (\ref{Peqn}) then defining
\be
\tilde x=\fft1{x}\,,\qquad \widetilde P(\tilde x)=\fft1{x}\, P(x)\,,
\ee
the function $\widetilde P(\tilde x)$ will solve the tilded equation
\be
\tilde x^2\, 
\del^2_{\tilde x}\, \widetilde P(\tilde x) + f(\tilde x, \tilde\delta_i)\, 
\widetilde P(\tilde x)=0\,.
\ee
Thus a solution $P(x)$ of the radial equation with charge 
parameters $\delta_i$ maps into a solution of the inverted radial equation
with charge parameters $\tilde\delta_i$ given by (\ref{invol}).

   Note that in the specialisation to pairwise-equal charges, such as 
$\delta_3=\delta_1$ and $\delta_4=\delta_2$, one has
\be
\tilde\delta_1=\tilde\delta_3=\delta_2\,,\qquad \tilde\delta_2=\tilde\delta_4=
\delta_1\,,
\ee
and so in this case $H(x,\delta_i)=H(\fft1{x},\delta_i)$
(since the function $H$ is symmetrical in the charge parameters). The constant
$\beta$ in this pairwise-equal case is given by
\be
\beta^2= -\fft{am}{\omega} + 2a^2(\cosh^2\delta_1\, \cosh^2\delta_2+
\sinh^2\delta_1\,\sinh^2\delta_2)\,.\label{pairelecbeta}
\ee
Thus in this case, and in its further specialisation to $\delta_1=\delta_2$
(all charges equal, i.e.~the Kerr-Newman solution studied by Couch and 
Torrence), the
inversion is an actual symmetry of the radial equation.

\subsection{Conformal inversion in the static limit}

   It is interesting to look at the limit where the extremal 4-charge
metric (\ref{4chargemet}) (with $m=a$) reduces to 
the metric of extremal 4-charge static black holes, first obtained as a  BPS black hole solution in \cite{cvyoII}.   Since the
physical charges in the extremal rotating metric 
are given by $q_i=a\sinh2\delta_i$, one must send the charge parameters
$\delta_i$ to infinity at the same time as sending the rotation parameter 
$a$ to zero, so as to hold the $q_i$ finite, and so one has
\be
q_i = \fft{a}{2}\, e^{2\delta_i}
\ee
fixed, and the metric becomes
\be
ds^2 = -\Big(\prod_i H_i\Big)^{-\ft12}\, dt^2 + 
\Big(\prod_i H_i\Big)^{\ft12}\, (dr^2 +
 r^2\, d\Omega^2)
\ee
in the extremal static limit, where $H_i=1+q_i\, r^{-1}$.  
After the inversion, and using (\ref{invol}), one has
\be
\tilde q_i = \fft{a}{2}\, e^{2\tilde\delta_i} =\fft{a}{2}\,
e^{\delta_1+\delta_2+\delta_3+\delta_4-2\delta_i}\,,
\ee
and hence 
\be
\tilde q_i = \fft{Q^2}{q_i}\,,\qquad Q^4\equiv \prod_i q_i\,.\label{tqqc}
\ee
This is precisely the relation between untilded and tilded charges that
was found in \cite{godgodpop} (and which we summarised in the Introduction) 
for the 4-charge static metrics and the
conformally-related  inverted metrics.  In that case, the transformation 
mapped the entire metric into another metric within the same 4-charge 
class. Thus the inversion symmetry (up to charge transformations) 
of the radial equation in the rotating case, which we exhibited above,
becomes an inversion symmetry (up to charge transformations) of the
entire metric in the static limit.

\section{Inversion Symmetry of Radial Equation for Pairwise Equal 
Dyonic STU Black Holes}
\label{sec:pairwisedyonic}

   The generalisation of the Couch-Torrence inversion symmetry of the
separated radial equation to the general case of the 8-charge dyonic
rotating extremal black holes of STU supergravity is rather complicated,
and we shall not present it here.  It becomes much more manageable in
special cases, such as the case with 4 electric charges, which we discussed
previously.  Another case that is relatively straightforward is when
the field strengths of STU supergravity are set equal in pairs, with
each of the two remaining independent fields carrying independent electric
and magnetic charges.

  From the paper \cite{chowcomp1} of Chow and Comp\`ere, after separating
variables in the massless Klein-Gordon equation for the 8-charge rotating
STU black holes, the radial equation takes the form
\bea
&&\del_r(R\, \del_r \, P(r)) + H\, P(r)=0\,,\nn\\
H&=& \fft{\omega^2\, W_r^2 - 2 a \omega k L_r 
+ a^2\, k^2}{R} + \lambda\,,\label{radH}
\eea
where $R=r^2-2m r + a^2-n^2$, the separation constant is $\lambda$, and
the factorised solutions are taken to have the form $\Psi=e^{-\im\omega t+
\im k\varphi}\, P(r)\, S(\theta)$.  (We follow \cite{chowcomp1} and 
use $k$ rather than $m$ for the azimuthal quantum number, since $m$ is
used here to denote the black-hole mass parameter.)  The constant
$n$ is given by
\be
n=- m\, \fft{\nu_1}{\nu_2}\,,
\ee
(this is the condition for the physical NUT charge $N=m\, \nu_1+ n\,\nu_2$
to be zero) and the functions $W_r$
and $L_r$ in the extremal case are given by
\bea
W_r^2 &=& \rho^4 + 4M \rho^2\,(\rho+m) + L_r^2\,,\nn\\
L_r &=& 2a^2\, \Big(\fft{\nu_2}{m}\, \rho + (\nu_2+2D)\Big)\,,
\eea
where we have defined 
\be
\rho=r-m\,,
\ee
and the quantities $\nu_1$, $\nu_2$ and $D$ are given, in the pairwise-equal 
case, by
\bea
\nu_1 &=& -\ft12 \sinh2\delta_1\, \sinh2\gamma_1 -
\ft12 \sinh2\delta_2\,\sinh2\gamma_2\,,\nn\\
\nu_2&=& \ft12\cosh2\delta_1\,\cosh 2\gamma_2 +
           \ft12\cosh2\delta_2\,\cosh 2\gamma_1\,,\nn\\
D&=& \ft14 (\cosh2\delta_1\, \cosh 2\gamma_2 -1)
           (\cosh2\delta_2\, \cosh 2\gamma_1 -1) \nn\\
&&+
      \ft14 \sinh 2\delta_1\, \sinh 2\delta_2\, \sinh 2\gamma_1\,
            \sinh 2\gamma_2\,.
\eea
The physical mass $M$, given in general by $M=m\mu_1+n\mu_2$ \cite{chowcomp1},
is given in the pairwise-equal case by $M=m \nu_2 - n\nu_1$ (since then  
$\mu_1=\nu_2$ and $\mu_2=-\nu_1$).
The four physical electric and magnetic charges $(Q_1,Q_2,P_1,P_2)$
carried by the two independent field strengths $F_1$ and $F_2$ are given
in terms of the boost parameters $(\delta_1,\delta_2,\gamma_1,\gamma_2)$
by \cite{chowcomp1}
\bea
Q_1&=& \ft12 m\, \sinh2\delta_1\, \cosh 2\gamma_2 + \ft12 n\,
\cosh2 \delta_1\, \sinh2\gamma_1\,,\nn\\
Q_2&=&\ft12 m\, \sinh2\delta_2\, \cosh 2\gamma_1 + \ft12 n\,
\cosh2 \delta_2\, \sinh2\gamma_2\,,\nn\\
P_1&=&\ft12 m\, \sinh2\gamma_1\, \cosh 2\delta_1 -\ft12 n\, 
   \cosh2\gamma_2\, \sinh 2\delta_1\,,\nn\\
P_2&=&\ft12 m\, \sinh2\gamma_2\, \cosh 2\delta_2 -\ft12 n\,
   \cosh2\gamma_1\, \sinh 2\delta_2\,.\label{QPcharges}
\eea

   Viewing $H$, defined in (\ref{radH}), as a function of $\rho$, we
find
\be
H(\rho) = \omega^2\, \Big[\rho^2 +\fft{\beta^4}{\rho^2}\Big] +
  \fft{4 \nu_2\, a^2\, \omega^2}{m}\, \Big[\rho + \fft{\beta^2}{\rho}\Big]
 + 4\nu_2\, a^2\, \omega^2 \Big(1+\fft{a^2\, \nu_2}{m^2}\Big) + \lambda\,,
\ee
where
\bea
\beta^2 &=& 2a^2\, (\nu_2+2D) - \fft{ak}{\omega}\,,\nn\\
&=& a^2\, (1+\cosh2\delta_1\,\cosh2\delta_2\,\cosh2\gamma_1\,\cosh2\gamma_2+
\sinh2\delta_1\,\sinh2\delta_2\,\sinh2\gamma_1\,\sinh2\gamma_2) \nn\\
&&-
   \fft{ak}{\omega}\,.\label{pairdyonbeta}
\eea
Thus the function $H$ has the inversion symmetry
\be
H(\rho) = H(\fft{\beta^2}{\rho})\,.
\ee
This implies an inversion symmetry of the radial equation, namely,
that if we define $\tilde \rho=\beta^2\, \rho^{-1}$, then the original
radial equation (\ref{radH}) implies the inverted equation
\be
\del_{\tilde\rho} (\tilde\rho^2\, \del_{\tilde\rho} \widetilde P(\tilde\rho))
+ H(\tilde\rho)\,\widetilde(\tilde\rho)=0\,,
\ee
where
\be
\widetilde P(\tilde\rho)=\fft{\rho}{\beta}\, P(\rho)=
\fft{\beta}{\tilde\rho}\, P(\fft{\beta}{\tilde\rho})\,.
\ee
Note that as in the case of the pairwise-equal specialisation of the
4 electric charge black holes discussed previously, the inversion is
an actual symmetry in this pairwise-equal dyonic charge case.  One can
easily verify that if the magnetic charges are set to zero, the inversion
here reduces to the previous pairwise-equal result, with the radius 
of inversion $\beta$ reducing from (\ref{pairdyonbeta}) to 
(\ref{pairelecbeta}).

\subsection{Conformal inversion symmetry for static pairwise-equal dyonic
black holes}

he static limit of the extremal rotating dyonic black holes with
pairwise-equal charges is achieved by sending $m$ (and hence $n$ and $a$)
to zero while sending the boost parameters to infinity, so as to keep
the physical charges in (\ref{QPcharges}) finite and non-zero.  This
can be done by sending
\be
m\rightarrow 4\bar m\, e^{-4\lambda}\,,\qquad 
\delta_i\rightarrow \delta_i +\lambda\,,\qquad
\gamma_i\rightarrow \gamma_i +\lambda\,,
\ee
and the taking the limit $\lambda\rightarrow\infty$.  The metric 
given in \cite{chowcomp1} becomes static, with
\bea
ds^2 &=& -\fft{r^2}{W}\, dt^2 + \fft{W}{r^2}\, (dr^2+ r^2\,d\Omega_2)\,,\nn\\
W &=& r^2 + 2M r + 2(\bar m^2+\bar n^2)\, 
 e^{2(\delta_1+\delta_2+\gamma_1 +\gamma_2}\,,\nn\\
\bar n &=& \bar m\,\fft{\cosh(\delta_1-\delta_2+\gamma_1-\gamma_2)}{
  \cosh(\delta_1-\delta_2-\gamma_1+\gamma_2)}\,,\nn\\
M&=& \bar m e^{\delta_1+\delta_2+\gamma_1+\gamma_2}\, 
\fft{\cosh^2(\delta_1-\delta_2+\gamma_1-\gamma_2)+
  \cosh^2(\delta_1-\delta_2-\gamma_1+\gamma_2)}{
\cosh(\delta_1-\delta_2-\gamma_1+\gamma_2)}\,.\label{staticmet}
\eea
The electric and magnetic charges in the static limit are given 
by\footnote{Note that $P_2=-P_1$ in the static limit.  Although this
might appear not to be a generic pairwise-equal dyonic configuration it
actually is, once one takes into account that there is an S-duality
of the pairwise-equal STU supergravity under which $(Q_1,P_1)$ and
$(Q_2,P_2)$ can both be rotated under
\bea
\begin{pmatrix} Q_i\\ P_i\end{pmatrix}\longrightarrow
\begin{pmatrix} \cos\theta &\sin\theta\\ -\sin\theta &\cos\theta\end{pmatrix}\,
\begin{pmatrix} Q_i\\ P_i \end{pmatrix}\,.
\eea
}
\bea
Q_1 &=& \fft{\bar m \, e^{2\delta_1}\, 
   [e^{2\delta_1}(e^{4\gamma_1}+e^{4\gamma_2})
  + 2 e^{2\delta_2+2\gamma_1+2\gamma_2}]}{2(e^{2\delta_1+2\gamma_2} +
e^{2\delta_2+2\gamma_1})}\,,\nn\\
Q_2 &=& \fft{\bar m\, e^{2\delta_2}\, 
[e^{2\delta_2}(e^{4\gamma_1}+e^{4\gamma_2})
  + 2 e^{2\delta_1+2\gamma_1+2\gamma_2}]}{2(e^{2\delta_1+2\gamma_2} +
e^{2\delta_2+2\gamma_1})}\,,\nn\\
P_1&=& \fft{\bar m\, e^{2\delta_1+2\delta_2}\, (e^{4\gamma_1}-e^{4\gamma_2})}{
2(e^{2\delta_1+2\gamma_2} +
e^{2\delta_2+2\gamma_1})}\,,\nn\\
P_2&=& \fft{\bar m\, e^{2\delta_1+2\delta_2}\, (e^{4\gamma_2}-e^{4\gamma_1})}{
2(e^{2\delta_1+2\gamma_2} +
e^{2\delta_2+2\gamma_1})}\,.
\eea

  A straightforward calculation shows that if we perform the inversion
\be
r=\fft{\alpha^2}{\tilde r}\,,\qquad \alpha^2 = 2(\bar m^2+\bar n^2)\, 
    e^{2(\delta_1+\delta_2+\gamma_1 +\gamma_2)}\,,
\ee
then $W=W(r)$ satisfies
\be
W(r)= \fft{\alpha^2}{\tilde r^2}\, W(\tilde r)\,,
\ee
and therefore the metric obeys the conformal inversion symmetry
\be
ds^2= \fft{\alpha^2}{\tilde r^2}\, d\tilde s^2\,,
\ee
where $d\tilde s^2$ is the same as the original metric $ds^2$ given in
(\ref{staticmet}), only
now written with $\tilde r$ in place of $r$.

\section{Conformal Inversion for 8-Charge Extremal Static STU Black Holes}

  The most general black hole solution in STU supergravity carries eight
independent charges, namely an electric and a magnetic charge for each
of the four electromagnetic fields.  The theory has an $SL(2,\R)^3$ global
symmetry, which can be thought of as $SL(2,\R)_1$ which is
an electric/magnetic S-duality; $SL(2,R)_2$, which acts on the 2-torus 
when one views the STU supergravity as the $T^2$ reduction of 
the 6-dimensional string; and $SL(2,R)_3$, which exchanges Kaluza-Klein
and winding fields. The $U(1)^3$ compact subgroup
of the $SL(2,\R)^3$ rotates the various charges, while keeping fixed the
asymptotic values of the scalar fields. Thus one can always employ the 
3-parameter $U(1)^3$ subgroup to reduce an arbitrary 8-charge black hole
to a canonical form in which there are only $5=8-3$ non-vanishing charge
parameters.

  This reduction to a 5-parameter canonical form was employed in 
\cite{cvettsey} in the case of the static BPS extremal STU black holes, in
order to construct a solution with 5 independent charge parameters.
They carry charges $(Q,P)$ of the form (after changing to a duality frame 
that matches the choice in our previous discussions) 
\be
(Q_1,0)\,,\qquad (Q_2,0)\,,\qquad (Q_3,p)\,,\qquad
   (Q_4,-p)\,.\label{cvts5charges}
\ee
For our present purposes it will be more convenient to re-introduce the
redundancy of the additional three parameters, so that we can present
the static extremal black holes in a symmetrical form with eight
independent charge parameters; four electric and four magnetic.  In order to
do this we shall construct explicitly the action of the 
$SL(2,\R)^3$ global symmetry on the STU supergravity fields and the charges,
and then make use of the $U(1)^3$ compact subgroup in order to derive
the general 8-charge solution from the 5-parameter solution presented in
\cite{cvettsey}.  Because these steps are a little involved, we relegate
them to appendices A and B.  

   The upshot from these calculations is that the
metric of the general 8-charge static extremal black hole is
\bea
ds^2 &=& -\fft{r^2}{\sqrt{V}}\, dt^2 + \fft{\sqrt{V}}{r^2}\, (dr^2 +
r^2\, d\Omega^2)\,,\label{8chstatic}\\
V&=& r^4 + \alpha\, r^3 + \beta\, r^2 + \gamma\, r + \Delta\,,\nn
\eea
where the coefficients $\alpha$, $\beta$, $\gamma$ and $\Delta$ are
obtained in eqns (\ref{genABCD2}) in appendix B, and which 
for convenience we reproduce here:
\bea
\alpha&=& \sqrt{\Big(\sum_i Q_i\Big)^2 +\Big(\sum_i P_i\Big)^2}\,,\nn\\
\beta&=& \sum_{i<j} (Q_i Q_j + P_i P_j)\,,\nn\\
\gamma&=& 
\fft1{\alpha}\, \Big[ 4\Delta +\ft12\beta^2 -
\ft12\sum_{i<j} (P_i^2 \,P_j^2 +Q_i^2 \,Q_j^2 + P_{ij}\, Q_{ij}) 
-3\prod_i P_i -3\prod_i Q_i \Big]\,,\nn\\
&&\nn\\
\Delta&=& \prod_i Q_i +\prod_i P_i +\ft12 \sum_{i<j} P_i Q_i P_j Q_j  
   -\ft14 \sum_i P_i^2 Q_i^2\,,\label{genABCD}
\eea
In the expression for $\gamma$, we have defined 
$P_{ij}= P_i P_j +(\prod_k P_k)/(P_i P_j)$ and $Q_{ij}=
Q_i Q_j + (\prod_k Q_k)/(Q_i Q_j)$ (so $P_{12}=P_1 P_2 + P_3 P_4$, etc.).

 If the radial coordinate of the 
metric (\ref{8chstatic}) is subjected to the inversion
\be
r\longrightarrow \tilde r=\fft{\sqrt{\Delta}}{r}\,,\label{inversion8}
\ee
then the conformally rescaled metric $d\tilde s^2$, defined by
\be
ds^2 = \fft{\sqrt{\Delta}}{\tilde r^2}\,d\tilde s^2\,, \qquad
d\tilde s^2= -\fft{\tilde r^2}{\sqrt{\widetilde V}}\, dt^2 +
\fft{\sqrt{\widetilde V}}{\tilde r^2}\, (d\tilde r^2 + \tilde r^2\, d\Omega^2)
\,,\label{8chtmet}
\ee
where $\widetilde V=\tilde r^4 + \tilde\alpha\, \tilde r^3 +
\tilde\beta\, \tilde r^2 + \tilde\gamma\, \tilde r + \widetilde\Delta$,
will be in the same class of black hole metrics provided that
there exists a mapping of the charges such that
\be
\tilde\alpha = \fft{\gamma}{\sqrt{\Delta}}\,,\qquad
\tilde\beta=\beta\,,\qquad \tilde\gamma= \alpha\, \sqrt{\Delta}\,,\qquad
\widetilde\Delta=\Delta\,.\label{invcons}
\ee
These four equations constitute the conditions that the eight mapped
charges $\widetilde Q_i$ and $\widetilde P_i$ must satisfy, if the 
conformally inverted 
metric is to be interpretable as again being contained within the 8-charge
family of static BPS extremal black holes. 

Note that the inversion of the inversion will give back
the original metric, up to a constant scale factor that we can always set
to unity by normalisation. The inversion will then be an involution, and the 
third equation in (\ref{invcons}) is automatically satisfied if the
first and the fourth are satisfied. Thus we can choose to view the four
conditions (\ref{invcons}) as instead being described by the first, second and
fourth equations in (\ref{invcons}), together with the normalisation 
condition for the inversion to be an involution.

  As one can easily verify, in the special case of the solution with just four 
electric charges, the expressions (\ref{genABCD}) reduce to
\be
\alpha=\sum_i Q_i\,,\qquad\beta=\sum_{i<j} Q_i Q_j\,,\qquad
\gamma= \sum_{i<j<k} Q_i Q_j Q_k\,,\qquad \Delta=\prod_i Q_i\,,
\ee
and the equations (\ref{invcons}) are indeed all satisfied if
the charges are mapped according to the rule (\ref{tqq}).

  For the general 8-charge case, we have not succeeded in finding an elegant
formula for a mapping of the charges that satisfies the conditions
in (\ref{invcons}).  However, we can see simply by counting the number of
conditions, and comparing with the number of unknowns, that we can expect
that a solution will always exist. If we include the normalisation choice that
the inverse of the inverse gives back exactly the original metric then, as
indicated in the discussion above, we have a total of four conditions to
be satisfied.  In the case of the solutions with four electric charges, 
this meant that the number of conditions was equal to the number of
unknowns (the four mapped charge parameters $\widetilde Q_i$, where it
was assumed also that the magnetic charges remained zero after the mapping).
Thus, in the 4-charge case it turned out that there was a unique solution
for the mapping of the charges, up to permutations, as given in (\ref{tqq}).  

   In the general 8-charge case we still have just four conditions to be
satisfied, but we now have eight unknowns (the mapped electric and magnetic
charges $\widetilde Q_i$ and $\widetilde P_i$).  Thus we can expect that
not merely will a solution exist for the mapped charges but that there
will in fact be a $4=8-4$ parameter family of possible solutions. This
perhaps accounts for the difficulty in finding an elegant solution in this
case for the mapping of the charges.  Unless there exists a criterion for
characterising what constitutes an ``elegant'' solution, it may not in
general be
possible to do more than just give the rule ``find the 4-parameter family
of solutions for the mapped charges $\widetilde Q_i$ and $\widetilde P_i$ 
that solve the four conditions (\ref{invcons}).''

  One way of giving a slightly more constructive description of the 
mapping of the charges is to consider the set of transformations under the
$SL(2,\R)^3$ global symmetry of the STU theory (see appendix A).  We know
that these transformations, implemented as in eqn (\ref{PQsl23}), leave the
quartic polynomial $\Delta$ defined in (\ref{genABCD}) invariant, 
which is precisely what is required by
the final equation in (\ref{invcons}).  The other quantities $\alpha$,
$\beta$ and $\gamma$ defined in (\ref{genABCD}) are not invariant under 
$SL(2,\R)^3$, although they are invariant under the $U(1)^3$ compact 
subgroup.  Thus we may view the 6-dimensional coset $SL(2,\R)^3/U(1)^3$
as parameterising candidate transformations of the 8 original untilded 
charges to give the 8 tilded charges that obey the remaining three conditions
in (\ref{invcons}).  

  A relatively simple possibility is to restrict the three $SL(2,\R)$ matrices
\be
\begin{pmatrix} a_i & b_i\cr c_i & d_i\end{pmatrix}
\ee
to be diagonal, 
\be
\begin{pmatrix} a_i & 0\cr 0 & \fft1{a_i}\end{pmatrix}\label{diagsl2r}
\ee
for $i=1$, 2 and 3.  This then leaves just the
three undetermined parameters $a_i$ to be solved, by requiring the remaining
equations (the first three) in eqn (\ref{invcons}) to be satisfied.  
Provided that there
exist real such solutions for the $a_i$, then the problem of finding the
mapping of charges under the conformal inversion is solved, in the sense
that it is reduced to solving three equations for three unknowns.

  As a first example, we may consider the 4-charge case where the 
magnetic charges are all zero, which was studied in \cite{godgodpop} and
is described in the introduction of the present paper. The mapped charges
$\widetilde Q_i$ are given in terms of the original charges $Q_i$
by (\ref{tqq}), and as can easily be checked from the formulae in appendix
A, they are produced by means of the $SL(2,\R)^3$ transformations
with the restricted form (\ref{diagsl2r}) where
\be
a_1= \sqrt{\fft{Q_1\, Q_2}{Q_3\, Q_4}}\,,\qquad
a_2= \sqrt{\fft{Q_1\, Q_3}{Q_2\, Q_4}}\,,\qquad
a_3= \sqrt{\fft{Q_1\, Q_4}{Q_2\, Q_3}}\,.
\ee

As another example, we may consider the special case of pairwise-equal
charges; without loss of generality we choose the case where the gauge
fields labelled 1 and 3 are set equal, and likewise the gauge fields 
labelled 2 and 4.  Thus we have
\be
Q_1=Q_3\,,\qquad Q_2=Q_4\,,\qquad P_3=P_1\,,\qquad P_4=P_2\,.
\ee
The $SL(2,\R)^3$ global symmetry of the full STU supergravity, described in 
appendix A, reduces to just the $SL(2,\R)_2$ symmetry.  
In this case the coefficients $\alpha$, $\beta$, $\gamma$ and $\Delta$ in
(\ref{genABCD}) become
\bea
\alpha &=&2\sqrt{(P_1+P_2)^2 + (Q_1+Q_2)^2}\,,\qquad
\beta= \ft14 \alpha^2 +2\sqrt{\Delta}\,,\nn\\
\gamma &=&\alpha\, \sqrt{\Delta}
\,,\qquad \Delta= (P_1 P_2 + Q_1 Q_2)^2\,,
\eea
and so the metric function $V$ becomes the perfect square
\be
V=\left(r^2 +\frac{\alpha\,r}{2} +\sqrt{\Delta}\right)^2\,.
\ee
After the inversion (\ref{inversion8}) and conformal scaling (\ref{8chtmet}),
the charge transformation conditions (\ref{invcons}) reduce simply to
\be
\tilde\alpha=\alpha\,,\qquad \widetilde\Delta=\Delta\,.\label{paircon}
\ee

  As we mentioned previously, the conformal inversion is actually a
symmetry in this pairwise-equal case, and correspondingly, 
as can be seen from (\ref{paircon}), one solution for the transformed 
(tilded) charges is simply to take them to be equal to the original
charges.  However, it is interesting to note that we can also find
other solutions to the conditions (\ref{paircon}) in which the charges 
are non-trivially transformed.  One way to do this is
by taking a diagonal $SL(2,\R)_2$ transformation with
\be
b_2=0\,,\qquad c_2=0\,,\qquad d_2=\fft1{a_2}\,.
\ee
 From the transformation (\ref{PQsl23}) we therefore find that 
eqns (\ref{paircon}) are satisfied if $a_2$ is chosen so that
\be
a_2^2 =  \fft{Q_1^2+P_2^2}{Q_2^2+P_1^2}\,.
\ee
The transformed charges are given explicitly by
\be
\widetilde Q_1 =\fft{Q_1}{a_2}\,,\qquad \widetilde Q_2 = a_2\, Q_2\,,\qquad
\widetilde P_1 = a_2\, P_1\,,\qquad
\widetilde P_2 =\fft{P_2}{a_2}\,.
\ee

  Another, inequivalent, 
way of solving (\ref{paircon}) in this pairwise-equal example
is to use instead a different 
non-compact $SL(2,\R)_2$ transformation, where
\bea
\begin{pmatrix} a_2 & b_2 \cr c_2 & d_2\end{pmatrix}=
\begin{pmatrix} \cosh\delta_2 &\sinh\delta_2 \cr 
                \sinh\delta_2 & \cosh\delta_2\end{pmatrix}\,.\label{sl2r2}
\eea
 From the transformation (\ref{PQsl23}) we now find that eqns (\ref{paircon})
are satisfied if
\be
\tanh\delta_2 = \fft{2(P_1 Q_1 - P_2 Q_2)}{P_1^2 + P_2^2+Q_1^2+Q_2^2}\,.
\ee
The transformations of the charges in this case correspond to a different
way of solving the constraint equations (\ref{paircon}).  This reflects the
fact that the constraints provide an undetermined system of equations.  In
this pairwise-equal specialisation, we have the two constraint equations
(\ref{paircon}) and the four unknowns $(\widetilde Q_1, \widetilde Q_2,
\widetilde P_1,\widetilde P_2)$.  The general solution would give a family
of transformed charges characterised by two continuous parameters.  We
have exhibited above two discrete members within this family, in 
addition to the ``trivial'' member where the charges are untransformed.

   In the 4-charge and the pairwise-equal examples above, it was possible
to present explicit expressions for solutions to the constraint equations.
As mentioned previously, this does not appear to be possible in the
general case with 8 independent charges.  For example, we can always
look for solutions for the transformed charges by considering the
subset of $SL(2,\R)^3$ transformations described by (\ref{diagsl2r}).  The
fourth constraint in (\ref{invcons}) is automatically satisfied because 
$\Delta$ is invariant under $SL(2,\R)^3$, and so
the remaining 3 constraints in (\ref{invcons}) will imply a discrete 
set of solutions for the 3 unknowns $(a_1,a_2,a_3)$.  The three equations
are polynomials in the $a_i$ parameters, but seemingly they are of too
high a degree to be explicitly solvable.  Of course for any specified
set of 8 original charges one can compute numerically the corresponding
$a_i$ parameter values that satisfy the constraints, and so in this sense
the problem is fully solvable.  There is, however, one obstacle that
can arise, namely, that it might happen that all of the solutions for the
$a_i$ parameters turn out to be complex.  We have looked at numerous 
examples of ``randomly chosen'' sets of original charges $(Q_1,Q_2,Q_3,Q_4,
P_1,P_2,P_3,P_4)$, and we find that sometimes real solutions for the
$a_i$ exist, while sometimes only complex solutions exist.  

  Another option would be to consider the subset of $SL(2,\R)^3$ 
transformations where each of the three $SL(2,\R)$ elements in the product 
is of the same form as
the $SL(2,\R)_2$ element (\ref{sl2r2}), giving three parameters 
$(\delta_1,\delta_2,\delta_3)$ in all.  Again, the remaining first three
equations in (\ref{invcons}) would give 3 equations for the 3 unknowns
$(\delta_1,\delta_2,\delta_3)$, guaranteeing that solutions would exist.
For a given initial choice for the 8 charges, it might be that while only
complex solutions arose for the three $a_i$ parameters in the previous 
construction, there could be real solutions for the three $\delta_i$ parameters
in the latter construction.  {\it Faute de mieux}, one could always, of course,
consider the most general possibility of simply viewing the four
conditions in (\ref{invcons}) as providing 4 equations for 8 unknowns.  It
is still not entirely obvious whether any purely real solutions for the
transformed charges must necessarily 
exist, but with such a large solution space 
it is perhaps likely.\footnote{The mapped charges should also, like the 
original ones, 
be non-negative, since otherwise
the conformally inverted metric $d\tilde s^2$ defined in (\ref{8chtmet})
would have naked singularities outside the horizon at $r=0$. In the examples
we found, when the charges come out to be real they are also non-negative.
It should be noted also that for the BPS extremal black holes to be regular,
without naked singularities, the quartic invariant $\Delta$ should be positive
(see, for example, \cite{larsenbps}).}  


\section{Inversion Symmetry of Radial Equation for 8-Charge Rotating 
Black Holes}

  We use the expressions and notation 
given in the paper \cite{chowcomp1} by
Chow and Comp\`ere.  It can be seen that the metric will be extremal
if $a^2=m^2+n^2$, and then the radial function $R$ will be given by
$R=(r-m)^2$.  We define a new radial coordinate $\rho=r-m$ that vanishes
on the horizon.  The separation of variables for a solution $\psi$
of $\square\psi=0$ is carried out in \cite{chowcomp1} by writing
$\psi=e^{-\im\omega t+\im k \varphi}\, \Phi_r(r)\,\Phi_u(u)$.  Their 
separated equations for $\Phi_r(r)$ and $\Phi_u(u)$ are presented in
eqn (9.17) of \cite{chowcomp1}:
\bea
&&\Phi_r^{-1}\, \fft{d}{dr}\Big(R\fft{d\Phi_r}{dr}\Big) +
\fft{\omega^2\,W_r^2 -2a\omega k\,L_r + a^2k^2}{R} +C_{cc} =0\,,\nn\\
&& \Phi_u^{-1}\, \fft{d}{du}\Big(U\fft{d\Phi_u}{du}\Big) -
\fft{\omega^2\,W_u^2 + 2a\omega k\,L_u + a^2k^2}{U} -C_{cc} =0\,.
\label{ccsep}
\eea
(We have renamed their separation constant as $C_{cc}$ rather than $C$,
since they already use $C$ for the quantity defined in eqn (5.5) of 
\cite{chowcomp1}.) The functions $W_r$, $L_r$, $W_u$ and $L_u$ 
are given in \cite{chowcomp1} (in fact $L_u=0$ when, as in our case, we
choose the physical NUT parameter $N$ to be zero).
The equations (\ref{ccsep}) are not quite in the form we
want, because the constant term in the potential in the equation for
$\Phi_u(u)$ has dependence on the charge parameters $\delta_i$ and 
$\gamma_i$.  Since we already know from our results in section 
\ref{sec:4chargerot} for the generalised Couch-Torrence symmetry of the
radial equation for the 4-charge solutions that the charge parameters 
are transformed in the inversion symmetry, we must ensure first in
the present 8-charge discussion that the angular equation for 
$\Phi_u$ should be independent of the charge transformations. This is
easily achieved, by exploiting the fact that there is always an arbitrariness
to shift the separation constant by an additive constant.  It  is 
straightforward to see from the definitions in \cite{chowcomp1} that
if we define $\bar u=u-n$ and $C_{cc}= -\lambda + 4\omega^2\,
[n\, (n\mu_1-m\mu_2) +(m^2+n^2)\,C]$, then angular equation becomes
\be
\Phi_u^{-1}\, \fft{d}{d\bar u}\Big(U\fft{d\Phi_u}{d\bar u}\Big) -
\omega^2\,U + \fft{a^2k^2}{U} +\lambda =0\,,
\ee
with $U=a^2-\bar u^2$, and this is completely independent of the
charge parameters. 

   The radial equation now takes the form
\be
\Phi_r^{-1}\, \fft{d}{d\rho}\Big(\rho^2\,\fft{d\Phi_r}{d\rho}\Big) +
H(\rho) -\lambda =0\,,
\ee
where $H(\rho)$ is given by
\bea
H(\rho) &=& \omega^2\, \Big\{ \rho^2 + \Big(2a^2\,(\nu_2+2D) -
      \fft{a\, k}{\omega}\Big)^2\, \rho^{-2}\Big\}\nn\\
&&+ 4\omega^2\,\Big\{M\, \rho + \fft{a^2\, \nu_2}{m}\, 
               \Big(2a^2\,(\nu_2+2D) -
      \fft{a\, k}{\omega}\Big)\,\rho^{-1}\Big\} \nn\\
&&+4\omega^2\, a^2\, (\mu_1+ C + \nu_1^2+\nu_2^2)\,,\label{Vfun}
\eea
and the various quantities $\nu_1$, $\nu_2$, $\mu_1$, $C$, $D$ and
$M$ are defined in \cite{chowcomp1}.

   We may now seek an inversion symmetry of the radial equation.  Thus we
transform to a new radial coordinate $\tilde\rho$ such that
\be
\rho=\fft{\beta^2}{\tilde\rho}\,,\label{rhoinversion}
\ee
where $\beta$ is a constant to be determined, and test to see whether
\be
H(\rho) =\widetilde H(\tilde\rho)\,,\label{Vtrans}
\ee
where the function $\widetilde H$ is the same in form as the function
$H$ given in (\ref{Vfun}), but using redefined charge 
parameters.\footnote{This, at least, is what we found for the 4-charge
rotating solutions in section \ref{sec:4chargerot}; the potential
in the radial equation was related under inversion to a potential of
the same form but with redefined charge parameters.}
Thus we have
\bea
H(\rho) &=& \omega^2\, \Big\{ \beta^4\,\tilde\rho^{-2} + 
\Big(2a^2\,(\nu_2+2D) -
      \fft{a\, k}{\omega}\Big)^2\, \beta^{-4}\, \tilde\rho^{2}\Big\}\nn\\
&&+ 4\omega^2\,\Big\{M\, \beta^2\, \tilde\rho^{-1} + \fft{a^2\, \nu_2}{m}\,
               \Big(2a^2\,(\nu_2+2D) -
      \fft{a\, k}{\omega}\Big)\,\beta^{-2} \,\tilde\rho\Big\} \nn\\
&&+4\omega^2\, a^2\, (\mu_1+ C + \nu_1^2+\nu_2^2)\,,\label{Vfuni}
\eea
Assuming that $\beta$ is universal, that is to say, that it is
invariant under the transformation of the charge parameters, we then
have, by comparing the various powers of $\rho$ or $\tilde \rho$ in
the proposed relation (\ref{Vtrans}) that
\bea
\rho^{-2}:\qquad && \beta^4 = 
   \Big(2a^2\,(\nu_2+2D) -\fft{a\, k}{\omega}\Big)^2\,,\nn\\
\rho^2:\qquad && \beta^4= 
  \Big(2\tilde a^2\,(\tilde\nu_2+2\widetilde D) 
             -\fft{\tilde a\, k}{\omega}\Big)^2\,,\nn\\
\rho^{-1}:\qquad && \widetilde M\, \beta^2 = 
\fft{a^2\, \nu_2}{m}\, \Big(2a^2\,(\nu_2+2D) -\fft{a\, k}{\omega}\Big)\,,\nn\\
\rho:\qquad &&
M\, \beta^2 =
\fft{\tilde a^2\, \tilde \nu_2}{\tilde m}\, 
\Big(2\tilde a^2\,(\tilde\nu_2+2\widetilde D) 
             -\fft{\tilde a\, k}{\omega}\Big)\,,\nn\\
\rho^0:\qquad && 
\mu_1+C +\nu_1^2 +\nu_2^2 = \tilde \mu_1 + \widetilde C + \tilde\nu_1^2 +
 \tilde\nu_2^2  \,.\label{fiveeqns}
\eea
From the first two equations in (\ref{fiveeqns}) we have
\be
\beta^2= 
2a^2\,(\nu_2+2D) -\fft{a\, k}{\omega}= 2\tilde a^2\,(\tilde\nu_2+2\widetilde D) 
             -\fft{\tilde a\, k}{\omega}\,.
\ee
Since we are assuming $\omega$ does not transform, and since this relation 
should hold for all frequencies $\omega$, it follows that
\be
\tilde a=a\,,\qquad \tilde\nu_2 + 2 \widetilde D= \nu_2 + 2 D\,.\label{nu2Dinv}
\ee
(Note that we are not making any assumption about $m$ or $n$ being invariant
under the transformation.)
The third and fourth equations in (\ref{fiveeqns}) then imply
\be
\widetilde M= \fft{a^2\, \nu_2}{m}\,,\qquad 
M= \fft{a^2\, \tilde\nu_2}{\tilde m}\,.\label{rhopm1}
\ee

   From \cite{chowcomp1}, $M=m\,\mu_1+n\,\mu_2$, and since we must set
$n=-m\, \nu_1/\nu_2$ so that the physical NUT charge $N=m\, \nu_1+
n\, \nu_2$ is zero, we have
\be
M=\fft{m}{\nu_2}\, (\mu_1\, \nu_2-\mu_2\,\nu_1)\,,
\ee
together with the transformed version where all quantities are tilded. 
Equations (\ref{rhopm1}) therefore imply
\be
a^2\, \nu_2\, \tilde\nu_2 = m\,\tilde m\, (\mu_1\, \nu_2-\mu_2\,\nu_2)=
m\,\tilde m\, (\tilde\mu_1\, \tilde\nu_2-\tilde\mu_2\,\tilde\nu_2)\,,
\ee
and hence
\be
\mu_1\, \nu_2-\mu_2\,\nu_2=\tilde\mu_1\, \tilde\nu_2-\tilde\mu_2\,\tilde\nu_2
\,.\label{munuinv}
\ee

  Collecting the results so far, we see from the last equation in 
(\ref{fiveeqns}), from (\ref{nu2Dinv}) and from (\ref{munuinv}) that the
three quantities
\be
X_1\equiv \mu_1+C +\nu_1^2 +\nu_2^2 \,,\qquad
X_2\equiv \nu_2+2D\,,\qquad X_3\equiv \mu_1\, \nu_2-\mu_2\,\nu_2
\label{threeinv}
\ee
should all be invariant under the transformation of the charge parameters
that accompanies the inversion (\ref{rhoinversion}).  A natural guess
for the inversion transformation, which would reduce to the known 
4-charge case $\gamma_i=0$ discussed in section \ref{sec:4chargerot} 
(and its duality partner where instead $\delta_i=0$), and would also reduce
to the known pairwise-equal dyonic case discussed in section 
\ref{sec:pairwisedyonic}, is to try
\be
\tilde\delta_i=-\delta_i+ \ft12 \sum_j\delta_j\,,\qquad
\tilde\gamma_i=-\gamma_i+ \ft12 \sum_j\gamma_j\,.\label{delgamtran}
\ee
One can in fact verify that the three quantities $X_1$, $X_2$ and $X_3$ 
defined in (\ref{threeinv}) are indeed invariant under (\ref{delgamtran}).
However, there is one further condition contained in the set of equations 
(\ref{fiveeqns}), since until now we just extracted the one condition 
(\ref{munuinv}) from the third and fourth equations in (\ref{fiveeqns}).
The remaining condition can be found by noting that the extremality
condition $a^2=m^2+n^2$ implies $a^2=m^2\, (1+\nu_1^2/\nu_2^2)$ (and its
tilded version), and using this in the third and fourth equations of
(\ref{fiveeqns}) leads to
\be
(\nu_1^2+\nu_2^2)(\tilde\nu_1^2+\tilde\nu_2^2)=X_3^2\,.
\ee
Straightforward calculation reveals that while this is indeed consistent
with (\ref{delgamtran}) in the 4-charge specialisation or in the
pairwise-equal specialisation (as it must be, since those cases were already
fully verified), it is not consistent in the general 8-charge case.  Thus
the transformation of the $\delta_i$ and $\gamma_i$ charge parameters in
the general case must be more complicated than the guess in
(\ref{delgamtran}).  However, since the number of conditions that must
be satisfied in order to achieve $H(\rho)=\widetilde H(\tilde\rho)$ 
is smaller than the number of unknown transformed charge parameters 
$(\tilde\delta_i,\tilde\gamma_i)$, we can conclude that it must be possible
to solve for such $(\tilde\delta_i,\tilde\gamma_i)$, even if we cannot
present the solution in a universal and elegant form.

\section{Concluding Remarks}

Studies of general rotating black holes in maximally supersymmetric ungauged 
supergravity theories,  (often referred to as STU black holes), 
revealed their numerous intriguing properties which often stem from, and  
provide an intriguing generalization of, properties of 
Kerr-Newman black holes in Einstein-Maxwell gravity.  Furthermore, 
the extremal black holes of that type are endowed with further enhanced 
symmetry properties, again generalising those of extremal Kerr-Newman 
black holes.   For example,  there has been substantial progress in recent 
studies  of the Aretakis charge both for extremal Reissner-Nordstr\"om  
black holes \cite{aretakis1,aretakis2} and  extremal Kerr black 
holes \cite{cvsa} as well as recent generalizations to 
extremal static 4-charge black holes \cite{godgodpop} and rotating 
ones \cite{cvposasa} in STU supergravity. Furthermore, in  
\cite{cvposasa} Aretakis charges for five-dimensional extremal STU black 
holes were derived, while generalisations  to related conserved charges 
of extremal static p-branes were given in \cite{cvposa}.

In this paper we focused principally on another type of symmetry, namely the 
Couch-Torrence symmetry of the radial part of the  massless 
Klein-Gordon equation for extremal rotating STU black holes.
This  symmetry \cite{coutor} was originally obtained for the radial equation 
of  the extremal Reissner-Nordstr\"om metric under the  conformal inversion 
transformation.
The generalisation of the  symmetry to  extremal rotating black holes, 
such as Kerr-Newman ones,  is possible due to the separability of the 
massless Klein-Gordon  equation in these backgrounds, where the 
inversion transformation of the radial equation depends not only on the 
black hole charge and the rotation 
parameters, but also on the mode eigenvalues for energy $\omega$  and 
azimuthal  angular momentum $m$.

In the present paper we further generalised this symmetry from the 
extremal static 4-charge  black holes \cite{godgodpop} to the extremal 
rotating  4-charge black holes in  STU supergravity.  We showed that in 
this case the radial equation for the separable massless Klein-Gordon 
equation indeed  exhibits an inversion symmetry, where again the inversion
radius depends not only on the black hole charges and rotation 
parameter, but also on the mode eigenvalues $\omega$ and $m$. However,  
unlike in the Kerr-Newman case,  after the  transformation the radial 
equation is different from original one, in the sense that 
it is the radial equation for a transformed set of the four charge parameters.  
This is a natural generalisation of the properties
of the static 4-charge STU extremal black holes, where a given black hole
is mapped under conformal inversion to another member of the 4-charge 
family with a transformed set of electric charges \cite{godgodpop}.

We also investigated  the  conformal inversion symmetry for the 
most general extremal BPS black holes in STU supergravity, 
specified by eight charges, both in the static and the rotating cases. 
For the static 8-charge case, which corresponds to BPS solutions,  
we showed that the entire  family of black holes maps into itself  
under conformal inversion.  However, unlike in the four-charge solutions, 
we were unable to give an elegant formula for the mapping of the 
eight charges under
conformal inversion. This is related to the fact that 
number of equations comprising the conditions for 
conformal inversion invariance is less than the number
of unknowns (the eight mapped charges), and thus there is no unique solution.   

In the case of extremal 8-charge rotating solutions,  
just as in the 4-charge case,  the  massless
Klein-Gordon equation is separable and  the behaviour of the 
radial wave equation under 
inversion can be investigating along the same lines.  However, just 
as in the static case, the inversion symmetry conditions are 
underdetermined, i.e. they do not fully constrain the mapped 
charges.\footnote{In the extremal  8-charge rotating case there is 
also another, non-BPS, branch. The extremal Kaluza-Klein dyon is a 
particular example of that type, which by itself does not possess the 
inversion symmetry, but should be mapped on onto an orbit of eight-charge 
non-BPS branch.}  We have provided a systematic procedure, which employs the
action of the coset generators of $\prod_{i=1}^3 SL(2,\R)_i/U(1)_i$
on the eight charges, in order to solve the constraints (\ref{invcons}) for
the mapped charges. 
Further investigations of these transformations are in order.

The generalisation of the Couch-Torrence symmetry to general extremal 
rotating black holes of STU supergravity demonstrates the 
existence of another symmetry of general asymptotically-flat black holes, 
extending the symmetries of  black holes in Einstein-Maxwell gravity.  
In spite of  the significantly more complicated field content and the 
structure of the Lagrangian,  which results in significantly more 
complicated metrics for the black holes, it is interesting that
the conformal inversion symmetry of 
the radial part of the massless Klein-Gordon equation 
persists in its generalised form.

 We expect that a generalisation of the Couch-Torrence symmetry may
persist also for general extremal black holes of STU supergravity in 
five dimensions. Again the
separability of the massless Klein-Gordon equation will play an important 
role. We defer further consideration of this case to future work.

We would like to conclude by emphasising that  our analysis focused on 
conformal inversion transformations for extremal BPS black holes in STU
supergravity for which the asymptotic values of the scalar fields 
were set to zero. It would, of course, be interesting to study 
conformal inversion transformations for such black holes with 
non-zero asymptotic values of the scalar fields. This explicit dependence 
on the asymptotic scalar fields is currently being investigated 
\cite{cvepopsah}. Such generalisations of the black hole solutions would 
in turn allow for generalisations of the Couch-Torrence transformations that 
could also involve scalar field transformations. Furthermore, one would be able to address  other types of inversion transformations, 
including those studied in \cite{borsduff2}.

 \vskip 1cm
\section*{Acknowledgments} We are grateful to Hadi Godazgar,  Mahdi 
Godazgar and Alejandro Satz  for helpful discussions.
The work of M.C. is supported in part by the DOE (HEP) Award DE-SC0013528, 
the Fay R. and Eugene L. Langberg Endowed Chair (M.C.) and the Slovenian 
Research Agency (ARRS No. P1-0306).  
The work of C.N.P. is supported in part by DOE 
grant DE-FG02-13ER42020.

\appendix

\section{Bosonic Lagrangian of STU Supergravity in Symmetrical Form}

  Here, we present the bosonic sector of the STU supergravity Lagrangian, in
a form where the four field strengths enter symmetrically.  
We take the Lagrangian as given in appendix B of \cite{10auth}, except
that the field strengths called $F^+$ in that paper will be called $F^-$
here, to match conveniently with the conventions
of Freedman and Van Proeyen \cite{FVP}.  
 
 In the notation of \cite{FVP}, but written in the language of 
differential forms, their eqn (4.66) becomes
\be
{\cal L}(F)= -\ft12 f_{AB}^R \, {*F^A}\wedge F^B + \ft{\im}{2}
f_{AB}^I\,{*\widetilde F^A}\wedge F^B\,,
\ee
where $f_{AB}^R$ and $f_{AB}^I$ denote the real and imaginary parts of
$f_{AB}$  (i.e $f_{AB}=f_{AB}^R + \im f_{AB}^I$).
Now $\widetilde F \equiv -\im\, {*F}$, and $*^2=-1$ in four-dimensional
spacetime when acting on 2-forms, so ${*\widetilde F}=\im\, F$ and hence
we have
\be
{\cal L}(F)= -\ft12 f_{AB}^R \, {*F^A}\wedge F^B - \ft{1}{2}
f_{AB}^I\,F^A\wedge F^B\,.\label{Flag}
\ee
Note that the field strengths $F^A$ are simply the exterior derivatives of 
potentials, $F^A=dA^A$.  Note also that $f_{AB}$ is symmetric in $A$ and $B$.

  As in \cite{FVP} we define $F^\pm=\ft12(F\pm \widetilde F)$, and hence
\be
F^\pm = \ft12(F \mp \im\, {*F})\,,\qquad {*F^\pm}= \pm\im F^\pm\,.
\label{Fpmdef}
\ee
Noting that for any 2-forms $X$ and $Y$ we have
\be
\ft12 X^{\mu\nu}\, Y_{\mu\nu}\, {*\oneone} ={*X}\wedge Y ={*Y}\wedge X\,,
\ee
we see that the wedge product of any self-dual 2-form with any anti-self-dual
2-form is zero:
\be
F^+\wedge F^-= -\im\, {*F^+}\wedge F^- =-\im\, {*F^-}\wedge F^+
= -F^-\wedge F^+ = - F^+\wedge F^-\,,
\ee
hence $F^+\wedge F^-=0$.  From (\ref{Fpmdef}) we have 
\be
F= F^+ + F^-\,,\qquad {*F}= \im (F^+ - F^-)\,.
\ee
 From the above, it follows that the Lagrangian (\ref{Flag}) can be written
as
\be
{\cal L}(F)= \ft{\im}{2}\, f_{AB}\, F^{-A}\wedge F^{-B} -
  \ft{\im}{2}\, \bar f_{AB}\, F^{+A}\wedge F^{+B}\,,
\ee
and hence 
\be
{\cal L}(F)= -\ft{1}{2}\, f_{AB}\, {*F^{-A}}\wedge F^{-B} -
  \ft{1}{2}\, \bar f_{AB}\, {*F^{+A}}\wedge F^{+B}\,.
\ee
In this form, we can compare with eqn (B.7) in \cite{10auth} (with the 
understanding that the roles of + and - superscripts on $F$ are exchanged
as mentioned previously),
and hence read off from (B.9) of \cite{10auth} 
that the matrix $f$, with components $f_{AB}$,
is given by
\be
f=  \fft1{W}\, \begin{pmatrix} e^{-\lambda_1} & e^{\varphi_1}\, \beta_1 &
e^{\varphi_2}\, \beta_2 & e^{\varphi_3}\, \beta_3 \cr
 e^{\varphi_1}\, \beta_1  & e^{-\lambda_2}\, \alpha_2\, \alpha_3 & 
  -e^{-\varphi_3}\, \alpha_3\, \beta_3 & -e^{-\varphi_2}\,\alpha_2\,\beta_2\cr
 e^{\varphi_2}\, \beta_2 & -e^{-\varphi_3}\, \alpha_3\, \beta_3 &
     e^{-\lambda_3}\, \alpha_1\, \alpha_3 & 
             -e^{-\varphi_1}\, \alpha_1\, \beta_1\cr
e^{\varphi_3}\, \beta_3 & -e^{-\varphi_2}\,\alpha_2\,\beta_2& 
   -e^{-\varphi_1}\, \alpha_1\, \beta_1 & 
        e^{-\lambda_4}\, \alpha_1\, \alpha_2 \end{pmatrix}\,,
\ee
where
\bea
\lambda_1&=& -\varphi_1-\varphi_2-\varphi_3\,,\quad 
\lambda_2= -\varphi_1+\varphi_2+\varphi_3\,,\quad
\lambda_3= \varphi_1-\varphi_2+\varphi_3\,,\quad
\lambda_4= \varphi_1+\varphi_2-\varphi_3\,,\nn\\
\alpha_i &=& 1 + e^{2\varphi_i}\, \chi_i^2\,,\nn\\
\beta_1 &=&e^{\varphi_2+\varphi_3}\, \chi_2\, \chi_3 + 
   \im\,e^{\varphi_1}\, \chi_1\,,\quad
\beta_2=e^{\varphi_1+\varphi_3}\, \chi_1\, \chi_3 +
   \im\,e^{\varphi_2}\, \chi_2\,,\quad
\beta_3=e^{\varphi_1+\varphi_2}\, \chi_1\, \chi_2 +
   \im\,e^{\varphi_3}\, \chi_3\,,\nn\\
W&=& 1 + \sum_i e^{2\varphi_i}\, \chi_i^2 - 
   2\im e^{\varphi_1+\varphi_2+\varphi_3}\, \chi_1\, \chi_2\, \chi_3\,.
\eea

  The field equations following from (\ref{Flag}) are $dG_A=0$, where
the 2-forms $G^A$ are read off from varying ${\cal L}(F)$ with respect to
$F^A$:
\be
\delta {\cal L}(F) = G_A\, \delta F^A = -f^R_{AB}\, {*F^B}  -
                   f^I_{AB}\, F^B\,,
\ee
and so
\be
  G_A = -f^R_{AB}\, {*F^B} -f^I_{AB}\, F^B\,.
\ee
It then follows that $G^\pm_A\equiv \ft12 (G_A \mp\im {*G_A})$ are given by
\be
G^-_A = \im\, f_{AB}\, F^{-B}\,,\qquad G^+_A= -\im\, f^*_{AB}\, F^{+B}\,.
\label{Gpmdef}
\ee
Note that the Bianchi identities $dF^A=0$ and the field equations $dG_A=0$
can be written as 
\be
d\Re(F^A)=0\,,\quad d\Re(G_A)=0\,,\qquad \hbox{or}\qquad
d\Im({*F^A})=0\,,\quad d\Im({*G_A})=0\,.
\ee
($\Re$ and $\Im$ denote the real and imaginary parts.)

   The Bianchi equations and equations of motion are invariant under
the transformations
\be
\begin{pmatrix} {F^\pm}\cr {G^\pm}\end{pmatrix}\longrightarrow 
\begin{pmatrix} {F^\pm}'\cr {G^\pm}'\end{pmatrix} ={\cal S}\,
  \begin{pmatrix} {F^\pm}\cr {G^\pm}\end{pmatrix} \,,\qquad \hbox{hence}\qquad
\begin{pmatrix} {F}\cr {G}\end{pmatrix}\longrightarrow
\begin{pmatrix} {F}'\cr {G}'\end{pmatrix} ={\cal S}\,
  \begin{pmatrix} {F}\cr {G}\end{pmatrix}\,,
\ee
where 
\be
{\cal S} =\begin{pmatrix} A&B\cr C&D\end{pmatrix}\,,\label{Sdef}
\ee
where $A$, $B$, $C$ and $D$ are real constant $4\times 4$ matrices, 
provided that the scalar fields transform appropriately:  We have
\be
{G^-}' = (C+\im D f)\, F^-= (C+\im D f)(A+\im B f)^{-1}\, {F^-}'\,.
\ee
Since the transformed fields must also obey (\ref{Gpmdef}), this implies
that the scalar matrix $f$ must transform according to \cite{FVP}
\be
\im f' =(C+\im Df)(A+\im Bf)^{-1}\,.\label{ftrans}
\ee
It is important that $f'$, like $f$, must be symmetric and so this implies
that the matrices $A$, $B$, $C$ and $D$ must obey the relations
\be
A^T C - C^T A=0\,,\qquad B^T D - D^T B=0\,,\qquad 
  A^T D - C^T B = \oneone\,.
\ee
These are precisely the conditions for the matrix ${\cal S}$ defined in 
(\ref{Sdef}) to be an element of $Sp(8,\R)$, obeying \cite{FVP}
\be
S^T\Omega S=\Omega\,,\qquad \hbox{where}\qquad \Omega=
\begin{pmatrix} 0&\oneone\cr -\oneone &0\end{pmatrix}\,.
\ee

   The scalar field Lagrangian is 
\be
{\cal L}(\varphi,\chi)= -\ft12\sum_i({*d}\varphi_i\wedge d\varphi_i
   + e^{2\varphi_i}\, {*d}\chi_i\wedge d\chi_i)\,,
\ee
and this is invariant under the $SL(2,\R)_1\times SL(2,\R)_2\times
  SL(2,\R)_3$, where the three $SL(2,\R)$ act in the standard way:
\be
\tau_i\longrightarrow \tau_i'=\fft{a_i\, \tau_i + b_i}{c_i\,\tau_i + d_i}\,,
\label{tautrans}
\ee
with $a_i\, d_i - b_i\, c_i =1$ for each $i$, and
\be
\tau_i = \chi_i + \im e^{-\varphi_i}\,.
\ee
Thus the entire STU supergravity theory is invariant under the intersection of
the $Sp(8,\R)$ symmetry of the field-strength sector and the
$SL(2,\R)_1\times SL(2,\R)_2\times SL(2,\R)_3$ symmetry of the scalar sector.
That is to say, the theory (at the level of the equations of motion) is
invariant under $SL(2,\R)_1\times SL(2,\R)_2\times SL(2,\R)_3$.

  To see how the field strengths and their duals transform under the
$SL(2,\R)_1\times SL(2,\R)_2\times SL(2,\R)_3$ symmetry, we just have to
work out the $A$, $B$, $C$ and $D$ matrices for which the transformation
(\ref{ftrans}) of the scalar field matrix matches with the 
$SL(2,\R)_1\times SL(2,\R)_2\times SL(2,\R)_3$ transformations 
(\ref{tautrans}).  After some algebra, we find
\crampest
\bea
A&=&\begin{pmatrix} a_1 a_2 a_3& -a_1 b_2 b_3 & -b_1 a_2 b_3 & -b_1 b_2 a_3\cr
       -a_1 c_2 c_3 & a_1 d_2 d_3 & b_1 c_2 d_3 & b_1 d_2 c_3\cr
    -c_1 a_2 c_3 & c_1 b_2 d_3 & d_1 a_2 d_3 & d_1 b_2 c_3 \cr
   -c_1 c_2 a_3 & c_1 d_2 b_3 & d_1 c_2 b_3 & d_1 d_2 a_3 \end{pmatrix}
\,,\quad
B=\begin{pmatrix} b_1 b_2 b_3 & -b_1 a_2 a_3 & -a_1 b_2 a_3 & -a_1 a_2 b_3\cr
   -b_1 d_2 d_3 & b_1 c_2 c_3 & a_1 d_2 c_3 & a_1 c_2 d_3 \cr
   -d_1 b_2 d_3 & d_1 a_2 c_3 & c_1 b_2 c_3 & c_1 a_2 d_3\cr
   -d_1 d_2 b_3 & d_1 c_2 a_3 & c_1 d_2 a_3 & c_1 c_2 b_3\end{pmatrix}\nn\\
&& \nn\\
C&=&\begin{pmatrix} c_1 c_2 c_3 & - c_1 d_2 d_3 &-d_1 c_2 d_3 &-d_1 d_2 c_3\cr
      -c_1 a_2 a_3 & c_1 b_2 b_3 & d_1 a_2 b_3 & d_1 b_2 a_3\cr
      -a_1 c_2 a_3 & a_1 d_2 b_3 & b_1 c_2 b_3 & b_1 d_2 a_3 \cr
      -a_1 a_2 c_3 & a_1 b_2 d_3 & b_1 a_2 d_3 & b_1 b_2 c_3
          \end{pmatrix}
\,,\quad
D=\begin{pmatrix} d_1 d_2 d_3 & -d_1 c_2 c_3 & -c_1 d_2 c_3 & -c_1 c_2 d_3\cr
  - d_1 b_2 b_3 & d_1 a_2 a_3 & c_1 b_2 a_3 & c_1 a_2 b_3 \cr
  -b_1 d_2 b_3 & b_1 c_2 a_3 & a_1 d_2 a_3 & a_1 c_2 b_3 \cr
  -b_1 b_2 d_3 & b_1 a_2 c_3 & a_1 b_2 c_3 & a_1 a_2 d_3\end{pmatrix}\nn
\eea
\uncramp
Note that $SL(2,\R)^3$ transformation matrix $S$ factorises into the 
(commuting) product of factors, $S=S_1 S_2 S_3$, where $S_i$ is the
transformation matrix for the $i$'th $SL(2,\R)$, which can be written
in the form (\ref{Sdef}) for very simple $4\times 4$ blocks $A_i$,
$B_i$, $C_i$ and $D_i$.  These can be read off from the general $A$,
$B$, $C$ and $D$ matrices above by setting $a_j=1$, $b_j=0$, $c_j=0$ and
$d_j=1$ for the two values of $j$ that are not equal to $i$. For example,
\crampest
\bea
A_1&=&\begin{pmatrix} a_1& 0& 0& 0\cr
       0 & a_1 & 0& 0\cr
   0& 0& d_1 & 0\cr
   0 & 0 & 0 & d_1 \end{pmatrix}
\,,\quad
B_1=\begin{pmatrix} 0 & -b_1 & 0& 0\cr
   -b_1 & 0& 0& 0 \cr
   0 & 0& 0& c_1\cr
   0 & 0 & c_1 & 0\end{pmatrix}\nn\\
&& \nn\\
C_1&=&\begin{pmatrix} 0 & - c_1 & 0&0\cr
      -c_1 & 0 & 0& 0\cr
      0 & 0 & 0& b_1  \cr
      0& 0& b_1 & 0
          \end{pmatrix}
\,,\quad
D_1=\begin{pmatrix} d_1 & 0& 0 & 0\cr
  0 & d_1 & 0& 0 \cr
  0 & 0 & a_1 & 0 \cr
  0 & 0 & 0 & a_1\end{pmatrix}\nn
\eea
\uncramp

  The magnetic and electric charges transform according to
\be
\begin{pmatrix} {\bf P}' \cr {\bf Q}'\end{pmatrix}=
  S\, \begin{pmatrix} {\bf P} \cr {\bf Q}\end{pmatrix}\,,\label{PQsl23}
\ee
where 
\be
{\bf P}= \begin{pmatrix} P_1 \cr P_2\cr P_3\cr P_4\end{pmatrix}\,,\qquad
{\bf Q} = \begin{pmatrix} Q_1\cr Q_2\cr Q_3\cr Q_4\end{pmatrix}\,.
\ee
The $U(1)^3$ compact subgroup of the full $SL(2,\R)^3$ symmetry group
corresponds to taking
\be
\begin{pmatrix} a_i & b_i\cr c_i & d_i\end{pmatrix} =
  \begin{pmatrix} \cos\theta_i &\sin\theta_i \cr
    -\sin\theta_i &\cos\theta_i\end{pmatrix}\,.\label{u13}
\ee

\section{8-Charge Static Black Hole}

  The 5-charge static black hole constructed in \cite{cvettsey} 
has metric given by
\be
ds^2 = -\fft{r^2}{\sqrt{V}}\, dt^2 + \frac{\sqrt{V}}{{r^2}}\, (dr^2 + r^2 d\Omega^2)\,, \label{exm}
\ee
where
\bea
V&=& (r+Q_1)(r+Q_2)(r+Q_3)(r+Q_4) -p^2 \,\Big[r + \ft12(Q_3+Q_4)\Big]^2\,,\nn\\
&=& r^4 + \alpha\, r^3 + \beta\, r^2 + \gamma\, r + \Delta\,, \label{V}
\eea
with
\bea
\alpha&=& \sum_i Q_i\,,\qquad \beta= \sum_{i<j} Q_i Q_j -p^2\,,\qquad
\gamma=\sum_{i<j<k} Q_i Q_j Q_k - p^2\, (Q_3+Q_4)\,,\nn\\
\Delta &=& \prod_i Q_i - \ft14 p^2\, (Q_3+Q_4)^2\,.\label{metABCD}
\eea
These 5-charge black holes correspond to restricting the charges 
$P_i$ and $Q_i$ 
of a general 8-charge static BPS extremal STU black hole by specialising  the
magnetic charges to 
\be
P_1=0\,,\qquad P_2=0\,,\qquad P_3=p\,,\qquad P_4=-p\,\label{5chspec}
\ee
as in (\ref{cvts5charges}).  To obtain the expressions for the
coefficients $\alpha$, $\beta$, $\gamma$ and $\Delta$ 
for the general case with 8 independent charges,
we may act on the restricted 5-charge solution with the 
$U(1)^3$ compact subgroup of the $SL(2,R)^3$ global
symmetry.  The eight parameters of the general solution then correspond to the
original five charge parameters plus the three parameters of 
the $U(1)^3$ rotations.
Conversely, we can determine the three $U(1)^3$ rotation angles
$\theta_1$, $\theta_2$ and $\theta_3$ such that the acting on a
general 8-charge configuration as in (\ref{PQsl23}) gives 
primed charges that are 
subject to the 5-charge specialisation in (\ref{5chspec}),
where we restrict to the $U(1)^3$ subgroup as defined in 
(\ref{u13}).  Defining
\be
\theta_\pm= \theta_2\pm\theta_3\,,
\ee
we find
\crampest
\bea
\tan \theta_+ &=& 
\fft{(P_1+P_2)-(Q_1+Q_2)\tan \theta_1}{(Q_3+Q_4)+(P_3+P_4)\tan\theta_1}\,,
\qquad
\tan\theta_-= 
\fft{(P_1-P_2)+(Q_1-Q_2)\tan \theta_1}{(Q_3-Q_4)-(P_3-P_4)\tan\theta_1}
\,,\\
&&\nn\\
\tan2\theta_1 &=& \fft{2(P_3+P_4)(Q_1+Q_2) -2(P_1+P_2)(Q_3+Q_4)}{
 (P_1+P_2+P_3+P_4)(P_1+P_2-P_3-P_4) +(Q_1+Q_2+Q_3+Q_4)(Q_1+Q_2-Q_3-Q_4)}\,.
\nn
\eea
\uncramp
It is also helpful to note that
\be
\tan(\theta_1+\theta_2+\theta_3)= \fft{P_1+P_2+P_3+P_4}{Q_1+Q_2+Q_3+Q_4}\,.
\ee

  After some algebra, we can now read off the general expressions for
the coefficients in (\ref{metABCD}) for the general 8-charge solutions.  
We find
\bea
\alpha&=& \sqrt{\Big(\sum_i Q_i\Big)^2 +\Big(\sum_i P_i\Big)^2}\,,\nn\\
\beta&=& \sum_{i<j} (Q_i Q_j + P_i P_j)\,,\nn\\
\gamma&=&
\fft1{\alpha}\, \Big[ 4\Delta +\ft12\beta^2 -
\ft12\sum_{i<j} (P_i^2 \,P_j^2 +Q_i^2 \,Q_j^2 + P_{ij}\, Q_{ij})
-3\prod_i P_i -3\prod_i Q_i \Big]\,,\nn\\
&&\nn\\
\Delta&=& \prod_i Q_i +\prod_i P_i +\ft12 \sum_{i<j} P_i Q_i P_j Q_j
   -\ft14 \sum_i P_i^2 Q_i^2\,,\label{genABCD2}
\eea
In the expression for $\gamma$, we have defined
$P_{ij}= P_i P_j +(\prod_k P_k)/(P_i P_j)$ and $Q_{ij}=
Q_i Q_j + (\prod_k Q_k)/(Q_i Q_j)$ (so $P_{12}=P_1 P_2 + P_3 P_4$, etc.).

 Note that $\Delta$ in (\ref{genABCD2}) is the standard quartic invariant, 
and it
is invariant under the full $SL(2,R)^3$ global symmetry group of the
STU theory.  The coefficients $\alpha$, $\beta$ and $\gamma$ 
in (\ref{genABCD2}) are
only invariant under the $U(1)^3$ subgroup.  This can be understood from the 
fact that the solutions have been chosen so that the scalar fields all go
to zero at infinity. Only the $U(1)^3$ subgroup of $SL(2,R)^3$
has the property of preserving this asymptotic condition on the scalars.

  If the radial coordinate of the static metric is subjected to the inversion
\be
r\longrightarrow \tilde r=\fft{\sqrt{\Delta}}{r}\,,
\ee
then the conformally rescaled metric $d\tilde s^2$, defined by
\be
ds^2 = \fft{\sqrt{\Delta}}{\tilde r^2} \, d{\tilde s}^2\,,\qquad
d\tilde s^2= -\fft{\tilde r^2}{\sqrt{\widetilde V}}\, dt^2 +
\fft{\sqrt{\widetilde V}}{\tilde r^2}\, (d\tilde r^2 + \tilde r^2\, d\Omega^2)
\,,
\ee
where $\widetilde V=\tilde r^4 + \tilde\alpha\, \tilde r^3 +
\tilde\beta\, \tilde r^2 + \tilde\gamma\, \tilde r + \widetilde\Delta$,
will be in the same class of black hole metrics provided that
there exists a mapping of the charges such that
\be
\tilde\alpha = \fft{\gamma}{\sqrt{\Delta}}\,,\qquad
\tilde\beta=\beta\,,\qquad \tilde\gamma= \alpha\, \sqrt{\Delta}\,,\qquad
\widetilde\Delta=\Delta\,.
\ee


\end{document}